\documentclass[aps,prm,twocolumn,superscriptaddress,nofootinbib,floatfix]{revtex4-2}

\usepackage{amsfonts}

\usepackage{graphicx}  
\usepackage{dcolumn}   
\usepackage{bm}        
\usepackage{amssymb}   

\usepackage{color}
\usepackage{amsmath}
\usepackage{soul}
\begin{document}

\title{Fit-Free Optical Determination of Electronic Thermalization  Time in Nematic Iron-Based Superconductors}

\author{Alexander Bartenev}
\affiliation{Department of Physics, University of Puerto Rico, Mayag\"uez, Puerto Rico 00681, USA}
\author{Roman Kolodka}
\affiliation{Department of Physics, University of Puerto Rico, Mayag\"uez, Puerto Rico 00681, USA}
\author{Ki-Tae Eom}
\affiliation{Department of Materials Science and Engineering, University of Wisconsin--Madison, Madison, WI 53706, USA}
\author{Jong-Hoon Kang}
\affiliation{Department of Materials Science and Engineering, University of Wisconsin--Madison, Madison, WI 53706, USA}
\author{Adri\'an R\'ua-Mel\'endez}
\affiliation{Department of Physics, University of Puerto Rico, Mayag\"uez, Puerto Rico 00681, USA}
\author{Jason Kawasaki}
\affiliation{Department of Materials Science and Engineering, University of Wisconsin--Madison, Madison, WI 53706, USA}
\author{Chang-Beom Eom}
\affiliation{Department of Materials Science and Engineering, University of Wisconsin--Madison, Madison, WI 53706, USA}
\author{Armando R\'ua}
\affiliation{Department of Physics, University of Puerto Rico, Mayag\"uez, Puerto Rico 00681, USA}
\author{Sergiy Lysenko}
\email{sergiy.lysenko@upr.edu}
\affiliation{Department of Physics, University of Puerto Rico, Mayag\"uez, Puerto Rico 00681, USA}

\begin{abstract}
	
We present a nematic response function model (NRFM) for fit-free direct extraction of the characteristic time of ultrafast electronic thermalization in iron-based superconductors, materials with electronic nematicity. By combining the NRFM for polarization-dependent pump--probe measurements of electronic nematic response with the two-temperature model (TTM) for sub-picosecond quasiparticle relaxation, we quantify the electronic thermalization timescales and their anisotropy. The nematic response function is modeled as the difference of normalized reflectivity signals, revealing a pronounced sub-picosecond extremum in signal evolution that directly yields the characteristic electronic thermalization time. This method demonstrates that the NRFM is consistent with TTM fits of transient optical response, yielding electronic thermalization time constants on the order of 110--230~fs for the FeSe$_{1-x}$Te$_x$ and Ba(Fe$_{0.92}$Co$_{0.08}$)$_2$As$_2$ thin films. The proposed approach can be applied to any material that exhibits electronic nematicity, providing a powerful tool for direct mapping of the relaxation time in nematic materials, avoiding complex experimental data-fitting procedures.

\end{abstract}

	\pacs{}
	\maketitle
	
\maketitle

\section{Introduction}

Fe-based superconductors (FBSs) show a rich interplay of electronic orders~\cite{Torchinsky2011,Mukasa2021,Lucarelli2010}, including antiferromagnetism~\cite{Su2009}, high-temperature superconductivity~\cite{Liu2012}, and an electronic nematic phase that breaks rotational tetragonal symmetry without long-range magnetic order~\cite{Harriger2011,Shimojima2019}. FBS compounds (pnictides~\cite{He2017,Nakajima2011} and chalcogenides~\cite{Bartlett2021,Chinotti2018, tanatar2016origin}) show strongly anisotropic normal-state properties, such as in-plane resistivity~\cite{He2017,Bartlett2021,tanatar2016origin} and optical conductivity~\cite{Nakajima2011,Chinotti2018}. These observations raise the question of whether unequal coupling of spin and orbital fluctuations to the electronic fluid produces the observed in-plane transport and optical anisotropies, and they motivate ultrafast investigations of electron--electron (\emph{e-e}) interactions and  electronic contributions to magnetoelastic and nematic channels~\cite{Konstantinova2019,Mansart2010,Patz2014}. However, developing a reliable approach to measure the effective electron thermalization time remains challenging.

A femtosecond pump--probe experiment can drive the system out of equilibrium and track its subsequent recovery~\cite{Torchinsky2011,Patz2014,torchinsky2010,Yang2018}: since electrons, lattice, and spins thermalize at very different rates, one can attribute features of the transient response to specific interactions based on their characteristic times. Specifically, electronically driven nematic order~\cite{Shimojima2019,Luo2017} is expected to respond on ultrafast timescales, revealing critical fluctuations that cannot be directly observed in static measurements. Moreover, the phenomenological two- and three-temperature models~\cite{Mansart2010, Patz2014, Rettig2013, Bartenev2023} can be used for FBS to obtain coupling strengths among electrons, lattice, and magnetoelastic degrees of freedom. 

In the optical pump–probe experiment, the time constant $\tau_{e}$ for the thermalization of the macroscopic electron subsystem reflects the change in the optical permittivity, averaged over $k$-space and orbitals. $\tau_{e}$ corresponds to the effective cooling of electrons via redistribution of spectral weight through all available channels and is sensitive to changes in collective correlations.

We also note that the information about electronic thermalization in FBSs can be obtained from ARPES measurements. ARPES can resolve $d_{xz}/d_{yz}$ band splitting and provide \emph{e-e} scattering rates \cite{Shimojima2019, Watson2015, Brouet2016}. However, the $\tau_{e}$ values obtained from optical experiments, in principle, differ somewhat from pure \emph{e-e} scattering times obtained from ARPES. Optical experiments provide $\tau_{e}$ values from $0.1$ up to $\sim0.5$~ps ~\cite{Patz2014, Stojchevska2012, Suzuki2017}, while \emph{e-e}~thermalization times from ARPES are usually shorter and strongly depend on momentum $k$ and energy.

In this work, we develop a methodology to obtain the hot-electron thermalization time in nematic materials, iron-based superconductors. It uses the nematic response function model (NRFM) for polarization-resolved pump--probe signals for direct extraction of electronic relaxation time constants along two orthogonal directions of materials with electronic nematicity. We start the analysis for the case of an infinitesimally short optical pulse. Then we generalize to a finite-pulse model, where the laser pulse has a certain temporal width. We compare the determined constants with the ones obtained using the two-temperature model (TTM), revealing a close match for the average scattering time constants, as well as adding insights about anisotropies of other quasiparticle (QP) relaxation processes. Here we study the nematic dynamics of FBSs below the superconducting transition temperature $T_{c}$. The proposed approach can be applied at any temperature, provided the material exhibits electronic nematicity.

\section{Nematic Response Function Model}

Transient reflectivity measurements provide insight into the carrier dynamics in a material by monitoring the time-dependent changes in reflectivity $\Delta R/R$. These changes are caused by interactions between QPs, specifically via the hot-electron thermalization followed by electron--phonon (\emph{e-ph}) scattering. For a two-component thermalization  process, the intrinsic material response can be modeled by a sum of exponential decays:
\begin{equation}
	\label{eq:two_exp}
	f(t) = A e^{-t/\tau_{e}} + B \left( 1 - e^{-t/\tau_{e-ph}} \right),
\end{equation}
where $A$ and $B$ are the amplitudes of the reflectivity change associated with \emph{e-e} and \emph{e-ph} relaxation mechanisms, respectively, and $\tau_{e}$ and $\tau_{e-ph}$ are their characteristic times.

The experimentally recorded signal is the intrinsic response convolved with the instrument response function (IRF)~\cite{dong2017pump, DielsRudolph}:
\begin{equation}
\left(\frac{\Delta R}{R}\right)(t)=(\mathrm{IRF}*f)(t)=\int_{-\infty}^{\infty}\mathrm{IRF}(t-t')\,f(t')\,dt'.
\label{eq:conv_irf_general}
\end{equation}
\emph{Infinitesimally Short Pulse Excitation.} In the limit of the infinitesimally short optical pulse, $\mathrm{IRF}(t)=\delta(t)$, Eq.~(\ref{eq:conv_irf_general}) reduces to $(\Delta R/R)(t)=f(t)$. We first derive the model in this limit, and then introduce corrections for a finite-time optical pulse.

\begin{figure*}
	\includegraphics[width=0.9\textwidth]{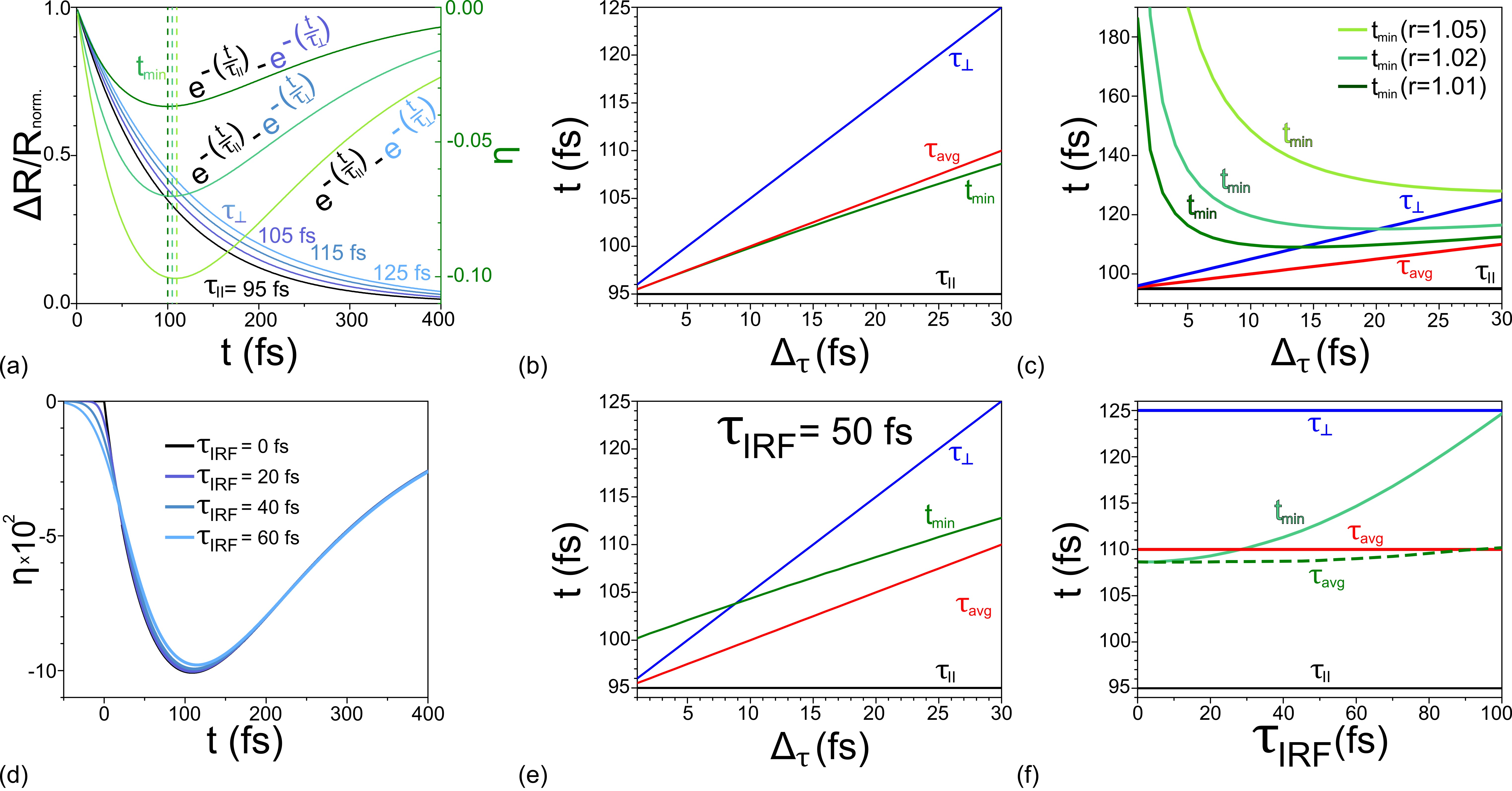}
	\caption{Nematic response function model. (a) Modeling of transient nematicity signal $\eta$ as a difference of normalized exponential decay functions $\exp(-{t/\tau_{\parallel, \perp}} )$ with characteristic relaxation times $\tau_{\parallel} = 95$~fs and $\tau_{\perp} =$ 105, 115, and 125~fs for the infinitesimally short optical excitation ($\tau_{\mathrm{IRF}}=0$~fs). 
		(b)~Minimum position $t_{\min}$ versus $\Delta_{\tau}$, compared to $\tau_{\perp}$, $\tau_{\parallel}$ and their average $\tau_{\mathrm{avg}}$ for $\tau_{\mathrm{IRF}}=0$~fs. 
		(c)~Divergence of $t_{\min}$ at small $\Delta_{\tau}$ when $\ln(r) \neq 0$ ($r=1.01, 1.02, 1.05$), and $\tau_{\mathrm{IRF}}=0$~fs.  
		(d)~Modeling of $\eta$ for finite-pulse excitation with $\tau_{\mathrm{IRF}}=0 - 60$~fs: difference of exponentials $\exp(-t/\tau_{\parallel,\perp})$ convolved with a Gaussian IRF ($\tau_{\parallel}=95$~fs, $\tau_{\perp}=125$~fs). 
		(e)~$t_{\min}$ versus $\Delta_{\tau}$ with a Gaussian IRF, compared to $\tau_{\perp}$, $\tau_{\parallel}$, and $\tau_{\mathrm{avg}}$ ($\tau_{\mathrm{IRF}}=50$~fs).
		(f)~Comparison between the modeled $\tau_{\mathrm{avg}}$ (solid line), the minimum position $t_{\min}$, and the $\tau_{\mathrm{avg}}$ (dashed line) calculated from Eq.~\eqref{eq:tauavg_IRF}.}
		\label{modelingNRFM}
\end{figure*}

Taking into account that the $\tau_{e} < \tau_{e-ph}$, at sufficiently short time scales, the term associated with \emph{e-ph} dynamics in Eq.~(\ref{eq:two_exp}) can be omitted: $B (1 - e^{-t/\tau_{e-ph}}) \approx 0$, and $\frac{\Delta R}{R} \approx A e^{-t/\tau_{e}} $ at $t < \tau_{e-ph}$. 

Measurements of the transient reflectivity at orthogonal polarizations $\perp$ and $\parallel$ enable observation of nematicity dynamics along orthogonal crystallographic directions. As will be shown below, the transient reflectivity signal for the two polarization channels can be modeled (at early times) by exponential decays,
$ \left(\frac{\Delta R}{R}\right)_{\perp} = A_{\perp} \exp\left(-{t/\tau_{\perp}}\right)$ and $\left(\frac{\Delta R}{R}\right)_{\parallel} = A_{\parallel} \exp\left(-{t/\tau_{\parallel}}\right)$, where $A_{\perp}$ and $A_{\parallel}$ are the amplitudes of the reflectivity change and $\tau_{\perp}$ and $\tau_{\parallel}$ are the characteristic electronic thermalization times for each polarization channel. The time difference $\Delta_{\tau}=\tau_{\perp}-\tau_{\parallel}$ is assumed to be relatively small.

The nematic response function can be constructed as
\begin{equation}
	\label{eq:nemresp1}
\tilde{\eta} = \left({\Delta R/R}\right)_{\parallel} - \left({\Delta R/R}\right)_{\perp}.
\end{equation}
By renormalizing it as $\eta = \dfrac{\tilde{\eta}}{A_{\perp}}$ with amplitude ratio $r=\dfrac{A_{\parallel}}{A_{\perp}}$, we obtain 
\begin{equation}
	\label{eq:nemresp2}
\eta = r \cdot \exp\left(-{t/\tau_{\parallel}}\right) - \exp\left(-{t/\tau_{\perp}}\right).
\end{equation}
This function has one extremum. For $\Delta_{\tau}>0$, this extremum is a minimum. Its position $t_{\min}$ can be obtained by solving the $\frac{d\eta}{dt}(t_{\min}) = 0.$
Assuming $\Delta_{\tau}\ll \tau_{\parallel}, \tau_{\perp}$,
\begin{equation}
		\label{eq:tmin1}
	t_{\min} \approx \tau_{\parallel} + \frac{\Delta_{\tau}}{2} + \ln(r)\left(\tau_{\parallel} + \frac{\tau_{\parallel}^2}{\Delta_{\tau}}\right).
\end{equation}

In case of equal amplitudes $A_{\perp} = A_{\parallel}$, $r = 1$ and $\ln(r) = 0$ we obtain
\begin{equation}
	\label{eq:tmin2}
	t_{\min} \approx \tau_{\parallel} + \frac{\Delta_{\tau}}{2} 
	= \frac{\tau_{\perp} + \tau_{\parallel}}{2} = \tau_{\mathrm{avg}}.
\end{equation}
This result indicates that the temporal position of the nematicity signal $\eta$ minimum (or maximum) corresponds to the average relaxation time $\tau_{\mathrm{avg}}$ of the material.

The nematic response function was modeled for various $\tau_{\perp}$ and $\tau_{\parallel}$ and is presented on Fig.~\ref{modelingNRFM}(a), where the maximum difference between $t_{\min}$ and $\tau_{\mathrm{avg}}$ is less than 2~fs ($\tau_{\mathrm{avg}}$=110~fs, $t_{\min}$=108.6~fs) for $\Delta_{\tau}$=30~fs with $\tau_{\parallel}$=95~fs and $\tau_{\perp}$=125~fs. The position of the minimum $t_{\min}$ of the nematic response function modeled for the $\tau_{\parallel} = 95$~fs and $\tau_{\perp} = 96-125$~fs is shown in Fig.~\ref{modelingNRFM}(b), indicating close match between actual $\tau_{\mathrm{avg}}$ and $t_{\min}$, in agreement with Eq.~(\ref{eq:tmin2}).

Furthermore, the value of the nematic response function in the minimum can be used to estimate the difference between electronic thermalization time constants for two directions. In case of $r = 1$, the expansion of the normalized nematic response function around $t_{\min}$ yields its amplitude $\eta_{\min}$ at the minimum position:
$\eta_{\min} \approx - \frac{\Delta_{\tau}}{e \cdot t_{\min}}.$ Therefore, the difference $\Delta_{\tau}$ between electronic thermalization times for two directions associated with maximal nematic response can be estimated as
\begin{equation}
	\label{eq:Deltatau}
	\Delta_{\tau} \approx -e \cdot \eta_{\min} \cdot t_{\min}.
\end{equation}
We note that, in Eq.~(\ref{eq:tmin1}), $t_{\min}$ diverges for $\Delta_{\tau} \to 0$ and $\ln(r) \neq 0$, as shown in Fig.~\ref{modelingNRFM}(c). Therefore, to obtain $t_{\min}$ with higher accuracy, it is important to renormalize the experimental data to enforce equal amplitudes $A_{\perp} = A_{\parallel}$, leading to $r = 1$ and $\ln(r) = 0$. We note that it is also convenient to rescale the amplitudes to unity, $A_{\perp} = A_{\parallel}=1$, without loss of generality. If the data are not normalized, the amplitude differences can cause errors between actual $t_{\min}$ and approximated $t_{\min} \approx \tau_{\mathrm{avg}}$ [Fig.~\ref{modelingNRFM}(c)]. The error increases with the decreasing $\Delta_{\tau}$. By combining~(\ref{eq:tmin2}) and~(\ref{eq:Deltatau}), it is possible to identify both $\tau_{\parallel}$ 
and $\tau_{\perp}$ as
\begin{equation}
		\label{eq:tau}
\tau_{\parallel,\perp} \approx  t_{\min}\left(1 \pm \tfrac{1}{2}e \cdot \eta_{\min}\right).
\end{equation}
Equations~(\ref{eq:Deltatau}) and~(\ref{eq:tau}), applied to extract electronic thermalization time from the transient reflectivity data will hereafter be referred to as the \textit{nematic response function model} (NRFM).

\emph{Finite-Pulse Excitation.}
For a finite duration of optical pulse, the experimentally recorded transient is the convolution of the intrinsic material response with the IRF~\cite{DielsRudolph}. For Gaussian pump and probe pulses of equal duration $\tau_p$ at the full width at half maximum (FWHM), the IRF is also Gaussian, with FWHM $\tau_{\rm IRF}=\sqrt{2}\,\tau_p$.

The renormalized ($A = 1$) intrinsic response is defined as
\begin{equation}
 f(t;\tau_{e})=
 \begin{cases}
  1, & t<0,\\[2pt]
  e^{-t/\tau_{e}}, & t\ge 0,
 \end{cases}
\label{eq:intrinsic_step}
\end{equation}
and the measured trace as $F(t;\tau_{e})=(g*f)(t)$.
The IRF can be approximated by a Gaussian of FWHM $\tau_{\mathrm{IRF}}$~\cite{DielsRudolph,TrinhEsposito2021}:
\begin{equation}
 g(t)=\frac{\kappa}{\sqrt{2\pi}\,\tau_{\mathrm{IRF}}}\exp\!\left[-\frac{\kappa^{2}\,t^{2}}{2\tau_{\mathrm{IRF}}^{2}}\right],
\label{eq:IRF_gauss}
\end{equation}
where $\kappa = 2\sqrt{2\ln 2}$. The convolution can be evaluated as
\begin{equation}
F(t;\tau_{e})=(\mathrm{g}*f)(t)=1-\Phi\!\left(\frac{\kappa\,t}{\tau_{\mathrm{IRF}}}\right)+S(t;\tau_{e},\tau_{\mathrm{IRF}}),
\label{eq:F_closed_main}
\end{equation}
where $\Phi(x)$ is the standard normal cumulative distribution function~\cite{CasellaBerger}
\begin{equation}
\Phi(x)=\frac{1}{2}\left[1+\operatorname{erf}\!\left(\frac{x}{\sqrt{2}}\right)\right]
\label{eq:Phi_def}
\end{equation}
and $S(t;\tau_{e},\tau_{\mathrm{IRF}})$
is
\begin{equation}
S(t;\tau_{e},\tau_{\mathrm{IRF}})=
\exp\!\left[\frac{\tau_{\mathrm{IRF}}^{2}}{2\kappa^{2}\tau_{e}^{2}}-\frac{t}{\tau_{e}}\right]
\Phi\!\left(\frac{\kappa\,t}{\tau_{\mathrm{IRF}}}-\frac{\tau_{\mathrm{IRF}}}{\kappa\,\tau_{e}}\right).
\label{eq:S_closed_main}
\end{equation}

The nematic response function is
\begin{equation}
\eta(t)=F(t;\tau_{\parallel})-F(t;\tau_{\perp}).
\label{eq:eta_F}
\end{equation}
Figure~\ref{modelingNRFM}(d) shows a comparison of $\eta(t)$ obtained for different IRFs by numerical simulation. Thus, the finite width of the optical pulse reduces the minimum amplitude $\eta_{\min}$ and shifts $t_{\min}(\Delta_{\tau})$ shown in Fig.~\ref{modelingNRFM}(b) for $\tau_{\mathrm{IRF}}=0$~fs to longer delays of several femtoseconds, as shown in Fig.~\ref{modelingNRFM}(e) for $\tau_{\mathrm{IRF}}=50$~fs. Although the difference between $\tau_{\mathrm{avg}}$ and $t_{\min}$ increases, it remains below 5~fs, indicating that for $\tau_{\mathrm{IRF}}=50$~fs the infinitesimally short pulse approximation remains valid.

For $|\Delta_{\tau}|\ll\tau_{\mathrm{avg}}$, $\eta(t)$ expanded around the mean relaxation time $\tau_{\mathrm{avg}}=(\tau_{\parallel}+\tau_{\perp})/2$ to obtain
\begin{equation}
\eta(t)\approx -\Delta_{\tau}\,\partial_{\tau_{\mathrm{avg}}}S(t;\tau_{\mathrm{avg}},\tau_{\mathrm{IRF}}).
\label{eq:eta_deltatau}
\end{equation}
Solving $d\eta/dt=0$ for $\kappa\,\tau_{\mathrm{avg}}/\tau_{\mathrm{IRF}}>3$ yields
\begin{equation}
 t_{\min}\approx \tau_{\mathrm{avg}}+\frac{\tau_{\mathrm{IRF}}^{2}}{\kappa^{2}\tau_{\mathrm{avg}}}
 \qquad \left(\frac{\kappa\,\tau_{\mathrm{avg}}}{\tau_{\mathrm{IRF}}}>3\right).
\label{eq:tmin_IRF_main}
\end{equation}
This leads to IRF-corrected NRFM relations (a detailed derivation is provided in the Supplementary Material):
\begin{equation}
\tau_{\mathrm{avg}} \approx t_{\min}-\frac{\tau_{\mathrm{IRF}}^{2}}{\kappa^2\,t_{\min}},
\label{eq:tauavg_IRF}
\end{equation}
\begin{equation}
\Delta_{\tau} \approx -e\,\eta_{\min}\,\tau_{\mathrm{avg}}\,
\exp\!\left(\frac{\tau_{\mathrm{IRF}}^{2}}{2\kappa^2\,\tau_{\mathrm{avg}}^{2}}\right),
\label{eq:Deltatau_IRF}
\end{equation}
\begin{equation}
\tau_{\parallel,\perp}\approx \tau_{\mathrm{avg}}\mp\frac{\Delta_{\tau}}{2}.
\label{eq:taus_IRF}
\end{equation}

Equation~\eqref{eq:tmin_IRF_main} shows that a finite time resolution $\tau_{\mathrm{IRF}}$ shifts the minimum to longer delays, with a quadratic dependence on $\tau_{\mathrm{IRF}}$. As shown in Fig.~\ref{modelingNRFM}(f), applying Eq.~\eqref{eq:tauavg_IRF} accounts for this shift and yields accurate $\tau_{\mathrm{avg}}$ even when $\tau_{\mathrm{avg}}\approx \tau_{\mathrm{IRF}}$.

\emph{Two-Temperature Model}.
The carrier and lattice dynamics can also be analyzed by the widely used TTM~\cite{Qiu1994,Norris2003}, 
which describes the temperature evolution of electrons $T_e$ and the lattice $T_L$ after a photoexcitation as
\begin{equation}
	\label{eq:TTM}
	C_e(T_e) \, \frac{\partial T_e(t)}{\partial t} 
	= -G \big(T_e(t) - T_L(t)\big) + S(\tau_{e}, t)
\end{equation}
\begin{equation}
	C_L(T_L) \frac{\partial T_L(t)}{\partial t} = G \big(T_e(t) - T_L(t)\big),
	\label{eq:TTM2}
\end{equation}
where $G$ is the \textit{e-ph} coupling constant, $C_e(T_e)$ and $C_L(T_L)$ are the electronic and lattice heat capacities per unit volume, respectively. The electronic heat capacity is a linear function of $T_e$: $C_e(T_e) = \gamma T_e$ and $\gamma$ is a Sommerfeld coefficient that depends on the density of states at the Fermi level~\cite{Smith2001}. The term $S(\tau_{e}, t)$ is a modified source term that accounts for non-instantaneous electron thermalization, defined by $\tau_{e}$~\cite{Naldo2020}:

\begin{equation}
	S(t) = \frac{0.94 (1 - R)}{(\tau_{p} + \tau_{e}) d} \, F \left( 1 - e^{-\tfrac{d}{\delta}} \right) 
	e^{\left[-2.77 \left(\frac{t - 2(\tau_{p} + \tau_{e})}{\tau_{p} + \tau_{e}}\right)^2 \right]}
	\label{eq:Source}
\end{equation}
where $F$ is the laser fluence, $d$ is the film thickness, and $\delta$ is the optical penetration depth. Transient reflectivity response can be modeled from the obtained electron and lattice temperatures:
\begin{equation}
	\frac{\Delta R}{R}(t) = a \Delta T_e(t) + b \Delta T_L(t)
	\label{eq:Reflectivity}
\end{equation}
where $a$ and $b$ are the fitting parameters, 
$\Delta T_e(t) = T_e(t) - T_e(0)$ and $\Delta T_L(t) = T_L(t) - T_L(0)$. Eq.~\eqref{eq:Reflectivity} can be used when the excitation remains reversible (i.e., below ablation/damage)~\cite{Mansart2010}. While the fluences used in this work (up to $8$~mJ/cm$^2$) are higher than in previous works~\cite{Konstantinova2019, Mansart2010}, they are still well below $25$ ~mJ/cm$^2$, whereat which the lattice temperature on a thin film surface can transiently exceed the melting temperature~\cite{suzuki2021ultrafast}. 

In this work, the two methodologies, NRFM and TTM, focus on different aspects: 
NRFM directly estimates $\tau_{e}$ from the nematic (anisotropic) signal at early times $t < \tau_{e-ph}$, whereas TTM fits the reflectivity signal, where the $\tau_{e}$ and $G$ are fitting parameters. By comparing the results, we can check for consistency and probe anisotropy in other relaxation channels, not addressed by TTM.

\section{Methods}

\emph{Samples}.
Epitaxial 40-nm-thick FeSe$_{0.8}$Te$_{0.2}$ and 190-nm-thick FeSe films were deposited on CaF$_2$ single-crystal substrates 
by pulsed laser deposition (PLD) at $10^{-6}$ Torr by Lambda Physik Compex 110 excimer laser with 
1.5 J/cm$^2$ fluence. During the deposition process, the substrate temperature was fixed at 573 K. 
Optimally doped 80-nm-thick Ba(Fe$_{1-x}$Co$_x$)$_2$As$_2$ ($x=0.08$) films were grown by PLD on 
(La, Sr)(Al, Ta)O$_3$ [LSAT] and LaAlO$_3$ [LAO] single-crystalline substrates with an epitaxial SrTiO$_3$ [STO] 
template buffer layer. STO template was etched by buffered hydrofluoric acid to achieve TiO$_2$ terminated surface.

\begin{figure}
	\includegraphics[width=\columnwidth]{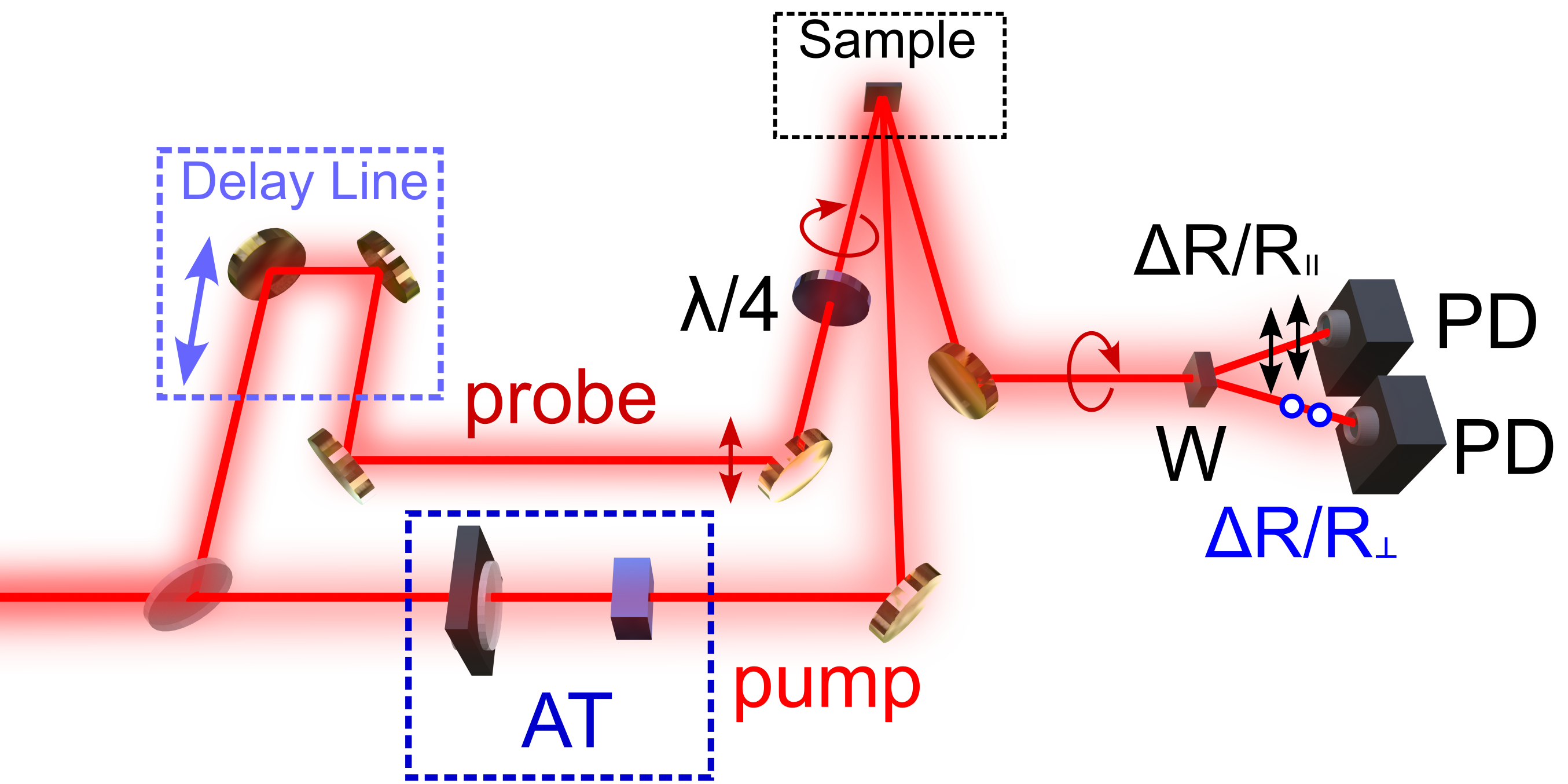}
	\caption{Schematic of the experimental setup for polarization-resolved pump--probe measurements. AT is pump intensity attenuator, $\lambda/4$ is quarter-wavelength plate, W is a Wollaston prism, and PD are photodetectors.}
	\label{fig:setup}
\end{figure}

\emph{Optical Measurements}.
In pump--probe measurements, we used the Spectra-Physics Ti:Sapphire femtosecond laser system with a pulse duration of $\tau_p=35$~fs, 800 nm central wavelength, and 1 kHz repetition rate.  The pump--probe experiments were carried out in reflection geometry (Fig.~\ref{fig:setup}), with the adjustable pump fluence up to 8~mJ/cm$^{2}$. The pump and probe pulses were overlapped on the sample surface, with substantially reduced intensity of probe pulses to avoid nonlinear interaction with the sample. The instrument time constant for 35~fs optical pulses is $\tau_{\mathrm{IRF}}\approx 50~\mathrm{fs}$. An electromechanical optical delay line controlled the time delay between pump and probe. The pump beam passed through the computer-controlled attenuator, enabling the precise control of the pump fluence. A circularly polarized probe was used for nematicity measurements. The circularity of the probe pulse was detected using a Wollaston prism. Samples were enclosed in an optical cryostat with a minimum possible temperature of 3 K.

\begin{figure*}
	\includegraphics[width=0.9\textwidth]{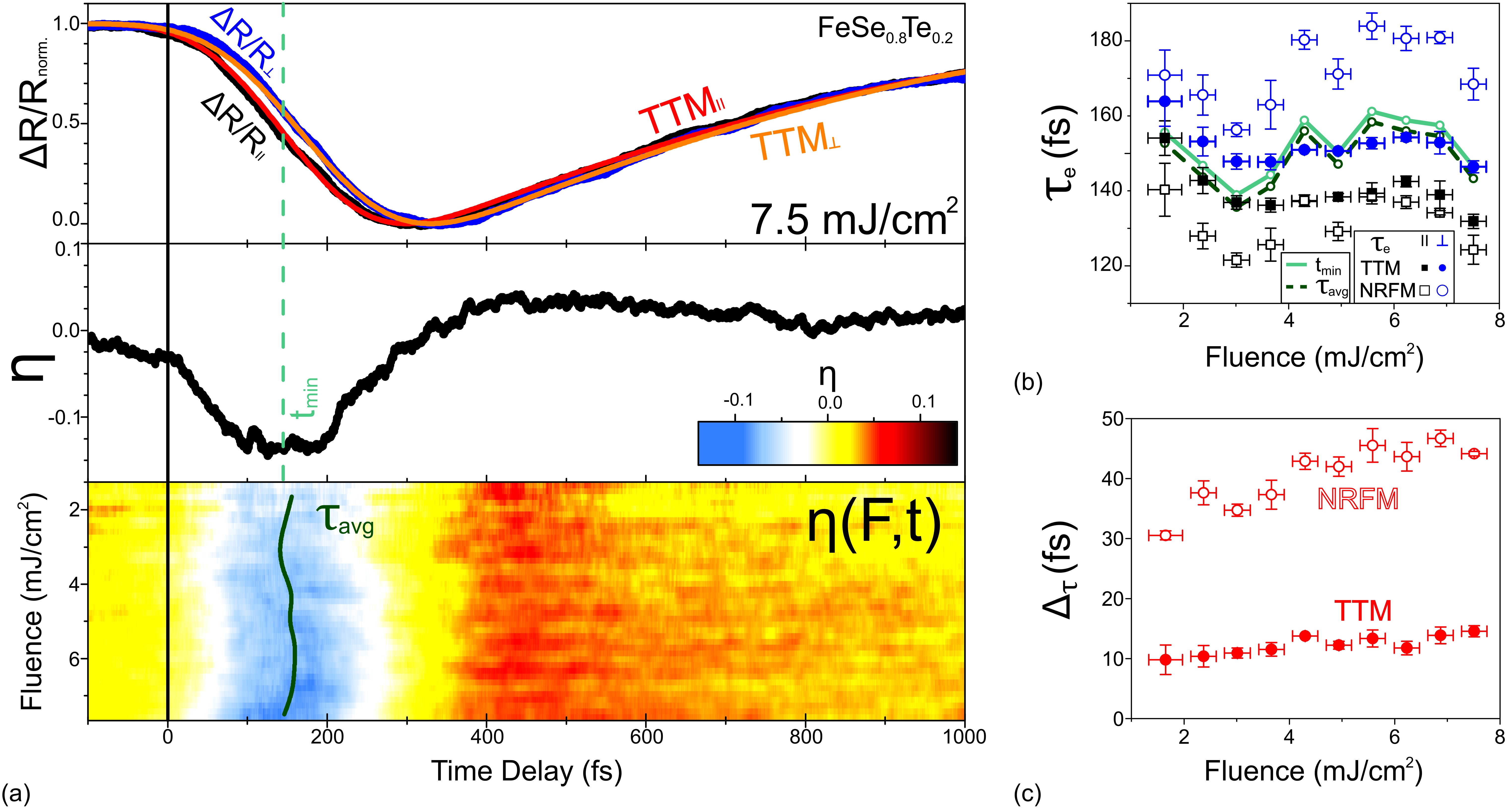}
	\caption{(a) Subpicosecond transient reflectivity of FeSe$_{0.8}$Te$_{0.2}$ at 8~K. Top panel: normalized transient reflectivity with superimposed TTM fits. Middle panel: nematic response function $\eta$ for transient reflectivity traces, shown in the top panel. Bottom panel: 2D map of the nematic response function $\eta$ versus photoexcitation fluence and time delay.  Solid line near the minimum $\eta_{\mathrm{min}}(F,t_{\mathrm{min}})$  corresponds to $\tau_{\mathrm{avg}}$ from Eq.~\eqref{eq:tauavg_IRF}. (b) Electronic thermalization time constants, obtained from TTM (solid) and from NRFM (hollow). Solid and dashed lines show $t_{\min}$ and $\tau_{\mathrm{avg}}$ from Eq.~\eqref{eq:tauavg_IRF}, respectively. (c) Differences $\Delta_{\tau}$, obtained from~(b).}
	\label{fig:Fig2}
\end{figure*}

Two polarization channels are recorded simultaneously using gains that provide equal amplitudes at the maximum of the transient signal. Gain balancing resulted in the relative scaling of the factor $r \equiv A_{\parallel}/A_{\perp}$,  as close to unity as possible. The remaining small deviation $r\neq 1$ due to sensitivity drift, differences in optical losses, or noise was eliminated by post-processing via a simple linear renormalization of a channel by a constant factor, making $A_{\parallel}=A_{\perp}$. This renormalization is equivalent to choosing a common measurement scale. It does not change the shape of the  normalized nematicity function $\eta(t)$ and, hence, the position of its extremum at $t_{\min}$. We note that the renormalization procedure does not introduce additional fitting parameters and does not affect the extracted characteristic time, defined by the time marker $t_{\min}$.

\section{Results and Discussion}

The ultrafast pump--probe measurements were performed for three samples: (i) FeSe$_{0.8}$Te$_{0.2}$ epitaxial film  at $T=8$ K, (ii) FeSe film at $T=3$ K, and (iii) Ba(Fe$_{0.92}$Co$_{0.08}$)$_2$As$_2$ at $T=8$ K.

Figure~\ref{fig:Fig2}(a) shows the transient reflectivity of FeSe$_{0.8}$Te$_{0.2}$. A sharp initial signal drop is attributed to hot-electron thermalization, followed by an increase attributed to the redistribution of excitation energy between electrons and lattice. The pronounced difference between normalized transient reflectivities along the two orthogonal directions results in photoinduced nematic response $\eta$(t). This function has a pronounced minimum at $t_{\min}\approx 150$ fs. The determination of $\tau_{\mathrm{avg}}$ here is based on the temporal position $t_{\min}$ of the extremum $\eta(t)$, but not on the absolute amplitude of the signal [Eq.~\eqref{eq:tauavg_IRF}]. The normalization of the channels as $A_{\parallel}=A_{\perp}=1$ with $r\rightarrow 1$ changes only the vertical scale of $\eta(t)$ and does not lead to a shift of $t_{\min}$.

\begin{figure*}
	\includegraphics[width=0.9\textwidth]{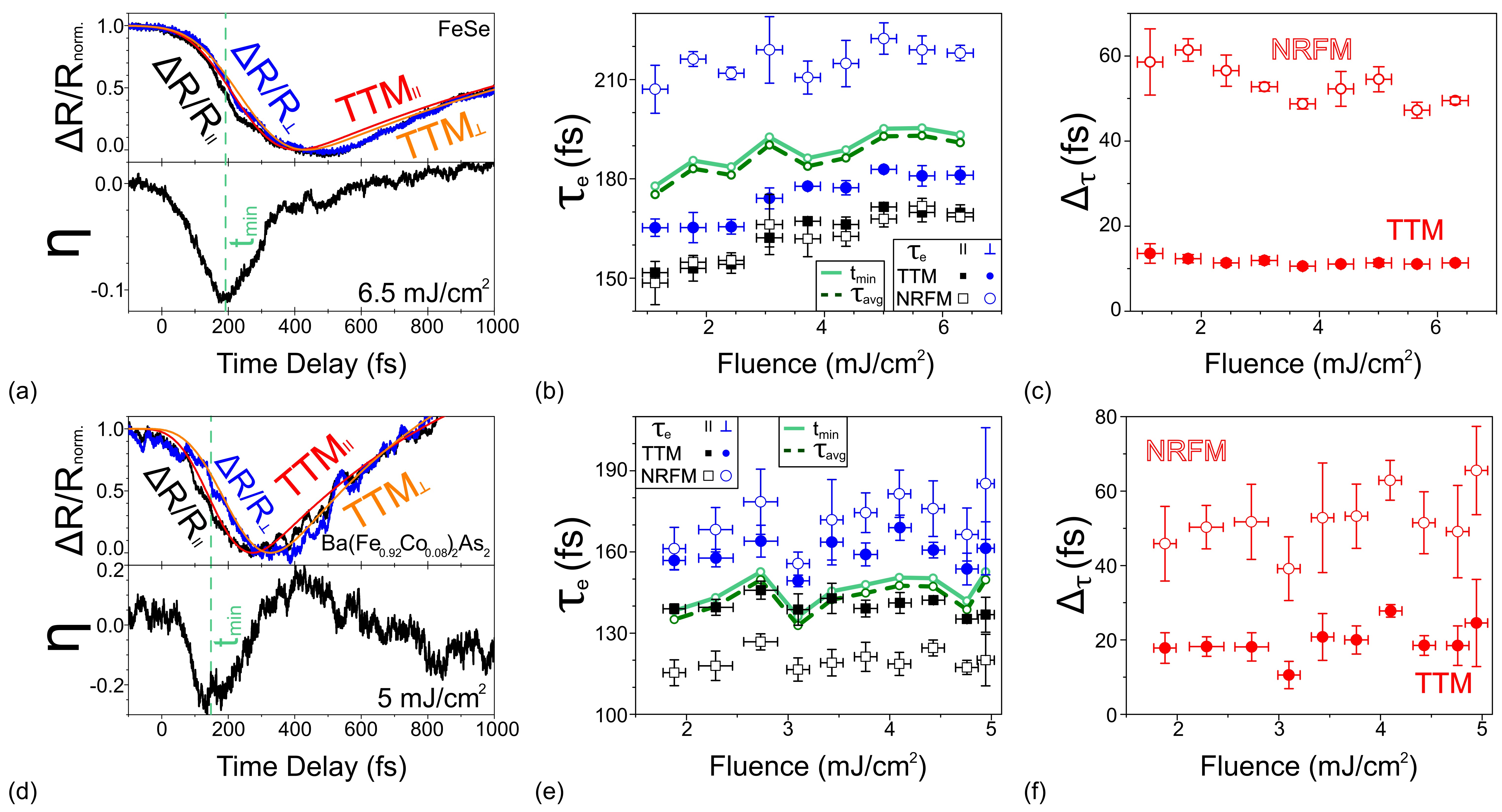}
	\caption{Subpicosecond transient dynamics of FeSe and Ba(Fe$_{0.92}$Co$_{0.08}$)$_2$As$_2$. 
		(a) Top panel: normalized transient reflectivity of FeSe thin film at 3~K with superimposed TTM fits. 
		Bottom panel: nematic response function $\eta$ for transient reflectivity traces, shown in the top panel. 
		(b) Electronic thermalization time constants, obtained from TTM (solid) and from NRFM (hollow), for FeSe thin film. Solid and dashed lines show $t_{\min}$ and the $\tau_{\mathrm{avg}}$ obtained from Eq.~\eqref{eq:tauavg_IRF}, respectively.
		(c) Differences  $\Delta_{\tau}$, obtained from (b). 
		(d) Top panel: normalized transient reflectivity of Ba(Fe$_{0.92}$Co$_{0.08}$)$_2$As$_2$ thin film at 8~K with superimposed TTM fits. Bottom panel: nematic response function $\eta$ for transient reflectivity traces, shown in the top panel. 
		(e) Electronic thermalization time constants, obtained from TTM (solid) and from NRFM (hollow), for Ba(Fe$_{0.92}$Co$_{0.08}$)$_2$As$_2$ thin film. Solid and dashed lines show $t_{\min}$ and $\tau_{\mathrm{avg}}$ from Eq.~\eqref{eq:tauavg_IRF}, respectively. 
		(f) Differences  $\Delta_{\tau}$, obtained from (e).}
	\label{fig:Fig3}
\end{figure*}

Since $t_{\min}$ is obtained from a position of local extremum $\eta(t)$ rather than from a multi-parameter fit, the reliability of $t_{\min}$ determination is high and is controlled by a few factors, such as (i) the signal-to-noise ratio near the extremum, (ii) baseline drift and normalization, (iii) the accuracy of the zero-time determination, and (iv) the width of the instrument response function $\mathrm{IRF}(t)$. The extraction of $\tau_\mathrm{avg}$ from $t_\mathrm{min}$ assumes that the analysis window is limited to early times when electron thermalization dominates ($t<\tau_{e\text{--}ph}$). For accurate anisotropy extraction, the amplitudes must be aligned, making  $r\rightarrow 1$. Otherwise, a possible amplitude mismatch can introduce a bias in $t_\mathrm{min}$ in the limit of small $\Delta\tau$ [see Fig.~\ref{modelingNRFM}(c)].

Figure~\ref{fig:Fig2}(b) shows that the TTM fitting yields electronic thermalization time constants $\tau_{e}$ that closely match the NRFM-derived time constants $\tau_{\mathrm{avg}}$ and $\tau_{\parallel,\perp}$ using Eqs.~\eqref{eq:tauavg_IRF} and~~\eqref{eq:taus_IRF},  respectively, revealing that both methods are capturing the same physical timescale. One can see the good agreement between NRFM (hollow points) and TTM (solid lines) -- the two sets of values are close to each other for all fluences. This consistency across a wide range of pump fluences validates the NRFM approach as a reliable method to obtain $\tau_{e}$ using the nematic signal alone. It also suggests that the modified TTM, which includes a finite $\tau_{e}$, is appropriately modeling the initial electronic excitation.

Figure~\ref{fig:Fig2}(c) shows the $\Delta_{\tau}$ values obtained using both models, TTM and NRFM [Eq.~\eqref{eq:Deltatau_IRF}]. The models predict the same fluence-dependent behavior of the differences $\Delta_{\tau}$ qualitatively. NRFM provides higher values of $\Delta_{\tau}$. The observed discrepancy can originate from the differences in approximations for each model.

The subpicosecond transient response of FeSe thin film is similar to observed dynamics of FeSe$_{0.8}$Te$_{0.2}$. Notably, the modified TTM fits the dynamics of FeSe thin film less accurately (the coefficient of determination $\mathfrak{R}^2$ of the fits $\geq 96\%$) than the FeSe$_{0.8}$Te$_{0.2}$ ($\mathfrak{R}^2 \geq 98\%$), as seen in Fig.~\ref{fig:Fig3}(a), where the modified TTM does not match the initial drop and subsequent rise as good as for FeSe$_{0.8}$Te$_{0.2}$ [Fig.~\ref{fig:Fig2}(a)]. It could be caused by a more pronounced nematic phase that Te doping suppresses: $T_s = 90$ K for undoped sample, temperature of structural transition reduces to $T_s = 60$ K for FeSe$_{0.8}$Te$_{0.2}$, structural transition completely vanishes for FeSe$_{0.5}$Te$_{0.5}$~\cite{Mukasa2021}. 

The electronic thermalization  constants $\tau_{e}$, obtained using both methods, show good match, with NRFM predicting higher $\tau_{\perp}$ [Fig.~\ref{fig:Fig3}(b)] and $\Delta_{\tau}$ [Fig.~\ref{fig:Fig3}(c)]. Notably, the divergence of $\Delta_{\tau}$ decreases with the fluence increase for FeSe, while increasing for FeSe$_{0.8}$Te$_{0.2}$. The origin of this divergence remains unclear and requires future investigation of the QP dynamics in these materials. Although the agreement between NRFM and TTM is less for FeSe than for FeSe$_{0.8}$Te$_{0.2}$ (both for $\tau_{e}$ and $\Delta_{\tau}$), the extracted mean value $\tau_{\mathrm{avg}}$ lies within one pulse width $\tau_p$ of both $\tau_{\perp}$ and $\tau_{\parallel}$.

The subpicosecond dynamics of optimally doped Ba(Fe$_{0.92}$Co$_{0.08}$)$_2$As$_2$ thin film [Fig.~\ref{fig:Fig3}(d)] is similar to the chalcogenides (FeSe and FeSe$_{0.8}$Te$_{0.2}$). For Ba(Fe$_{0.92}$Co$_{0.08}$)$_2$As$_2$ [Fig.~\ref{fig:Fig3}(e)], the electronic thermalization time constant $\tau_{e}$ does not show a clear fluence dependence  and both models yield comparable values. The $\tau_{e}$ values differ by only $\sim\tau_p$, i.e., within the experimental time resolution. The extracted $\Delta_{\tau}$ values [Fig.~\ref{fig:Fig3}(f)] differ more between the models, but the discrepancy is still on the order of the pulse duration.

Overall, the mean values $\tau_{\mathrm{avg}}$ obtained from NRFM are consistent with the $\tau_{e}$ values from TTM within the pulse duration and in agreement with typical values obtained for FBSs ~\cite{Patz2014, Stojchevska2012, Suzuki2017}. Moreover, the IRF correction to $\tau_{\mathrm{avg}}$ ($\sim 3$~fs in this work) is much smaller than the pulse duration and can be neglected. Thus, the NRFM approximation of infinitesimally short optical pulse excitation with $\tau_{\mathrm{IRF}}=0$~fs, where $\tau_{\mathrm{avg}} \approx t_{\min}$, is sufficient. Additionally, these approximation errors are systematic and therefore do not affect the predicted relative trends. To reduce uncertainties associated with the finite IRF, one can use shorter optical pulses.

According to Gurzhi's theory ~\cite{Maslov1, Gurzi1}, one can expect a noticeable decrease in $\tau_{e}$ with increasing pump fluence. However,  $\tau_{e}$ shows a weak dependence on the excitation level for FBSs in the superconducting regime: a slight decrease in $\tau_{e}$ vs pump level is observed only for FeSe$_{0.8}$Te$_{0.2}$ [Fig.~\ref{fig:Fig2}(b)], while for  Ba(Fe$_{0.92}$Co$_{0.08}$)$_2$As$_2$  almost no dependence on the pump fluence was detected [Fig.~\ref{fig:Fig3}(e)]. An unexpected growth in $\tau_{e}$ with increasing pump fluence is noteworthy for FeSe [Fig.~\ref{fig:Fig3}(b)]. Such unusual behaviors are associated with specific photoinduced dynamics of QPs and Cooper pair breakdown on sub-ps time scales in superconducting regime. We also note that the preliminary studies of $\tau_{e}$ showed that the trend of $\tau_{e}$ vs pump fluence changes above $T_{c}$. Therefore, temperature-dependent studies of $\tau_{e}(F)$  are promising for understanding the nature of superconductivity in FBSs.

The times $\tau_{\mathrm{avg}}$, $\tau_{\perp}$, and $\tau_{\parallel}$ extracted by NRFM should be understood here as the characteristic times of electron thermalization of the quasi-equilibrium electron subsystem, observed through the ultrafast evolution of the nematic optical response $\eta$(t), rather than as the direct microscopic time of single \emph{e-e} scattering. Extracted time shows how quickly the electron subsystem, after photoexcitation, relaxes to a quasi-thermal distribution and can include the contribution of several fast processes (e.g., energy redistribution within the electron band, relaxation of the distribution anisotropy, and ultrafast nematic channels). This time is somewhat different from the \emph{e-e} scattering time determined by ARPES, which pertains to a more microscopic time scale of electron state broadening, and depends on the specific definition of the observable. Therefore, here $\tau_{\mathrm{avg}}$, $\tau_{\perp}$, and $\tau_{\parallel}$ are integral measures of the electron thermalization rate, directly accessible from optical experiments.

\section{Conclusion}

In summary, a robust methodology based on polarization-resolved transient reflectivity measurements was developed for direct extraction of ultrafast electronic thermalization times in iron-based superconductors with electronic nematicity. The agreement between the time constants obtained by NRFM and independent TTM analysis confirms the consistency of the approach. The electronic thermalization time constants obtained for Ba(Fe$_{0.92}$Co$_{0.08}$)$_2$As$_2$ and FeSe$_{1-x}$Te$_x$ by using NRFM show good agreement with time constants obtained from the widely used TTM.

An important advantage of NRFM is that the electronic thermalization time is extracted from a stable time marker $t_{min}$ for the extremum of the nematicity function $\eta$. This makes the procedure insensitive to the choice of fitting parameters as compared to global multi-exponential or TTM approaches. Therefore, the NRFM method is well suited for high-throughput analysis of relaxation trends as a function of temperature, doping, or fluence. The proposed nematic response function model offers a quantitative probe of anisotropic electron dynamics not only for iron-based superconductors but also for a broad class of correlated materials with electronic nematicity, enabling quantification of the photoinduced dynamics of the nematic state.                                                                                  

\section*{Supplementary material}
See the supplementary material for corrections to the nematic response function model that account for the finite time resolution of the experiment.

\begin{acknowledgments}
The authors gratefully acknowledge support from the U.S. Army Research Office, accomplished under Grant Number W911NF-25-1-0122, and from the National Science Foundation, Awards \#2425113 and \#1905691. CBE acknowledges support for this research through a Vannevar Bush Faculty Fellowship (ONR N00014-20-1-2844), and the Gordon and Betty Moore Foundation's EPiQS Initiative, Grant GBMF9065. Thin film synthesis and transport measurements at the University of Wisconsin--Madison were supported by the U.S. Department of Energy (DOE), Office of Science, Office of Basic Energy Sciences (BES), under award number DE-FG02-06ER46327. The authors gratefully acknowledge partial support of this research by NSF through the University of Wisconsin Materials Research Science and Engineering Center (DMR-2309000).
\end{acknowledgments}

%

\clearpage
\onecolumngrid
\appendix


\section*{Supplementary Material}

\section*{Instrument Response Function Correction in the Nematic Response Function Model}

\subsection*{1. Definitions}

We model the intrinsic material response as
\begin{equation}
	f(t;\tau)=
	\begin{cases}
		1, & t<0,\\[2pt]
		e^{-t/\tau}, & t\ge 0,
	\end{cases}
	\label{eq:1}
\end{equation}
which can be written using the Heaviside step function $H(t)$ as
\begin{equation}
	f(t;\tau)=1-H(t)\bigl(1-e^{-t/\tau}\bigr)=1-H(t)+H(t)e^{-t/\tau}.
	\label{eq:2}
\end{equation}

The instrument response function (IRF) in a pump--probe measurement is the cross-correlation of the pump and probe pulses which characterizes the time resolution of the experiment~\cite{DielsRudolph2,SalehTeich}.
For Gaussian pump and probe pulses of equal duration $\tau_p$ at the full width at half maximum (FWHM), the IRF is also Gaussian, with FWHM $\tau_{\rm IRF}=\sqrt{2}\,\tau_p$. The corresponding root-mean-square (RMS) width $\sigma$ is~\cite{DielsRudolph2,SalehTeich}:
\begin{equation}
	\sigma \equiv \frac{\tau_{\mathrm{IRF}}}{2\sqrt{2\ln 2}}
	=\frac{\tau_{\mathrm{IRF}}}{\kappa},
	\qquad \kappa\equiv 2\sqrt{2\ln 2}.
	\label{eq:3}
\end{equation}

The normalized Gaussian IRF is then
\begin{equation}
	g(t)=\frac{1}{\sqrt{2\pi}\sigma}\exp\!\left(-\frac{t^{2}}{2\sigma^{2}}\right),
	\qquad
	\int_{-\infty}^{\infty} g(t)\,dt=1,
	\label{eq:4}
\end{equation}

The measured (convolved) signal is
\begin{equation}
	F(t;\tau)=(g*f)(t)=\int_{-\infty}^{\infty} g(t-u)\,f(u;\tau)\,du.
	\label{eq:5}
\end{equation}

\subsection*{2. Explicit form of the convolved single-channel trace}

Inserting Eq.~\eqref{eq:2} into Eq.~\eqref{eq:5} yields
\begin{equation}
	F(t;\tau)= (g*1)(t) - (g*H)(t) + \bigl(g*(H e^{-t/\tau})\bigr)(t).
	\label{eq:6}
\end{equation}
Because $g$ is normalized, $(g*1)(t)=1$, hence
\begin{equation}
	F(t;\tau)=1-(g*H)(t)+S(t;\tau,\sigma),
	\label{eq:7}
\end{equation}
where we define
\begin{equation}
	S(t;\tau,\sigma)\equiv \bigl(g*(H e^{-t/\tau})\bigr)(t)
	=\int_{0}^{\infty} g(t-u)\,e^{-u/\tau}\,du.
	\label{eq:8}
\end{equation}

The convolution with the step function is
\begin{equation}
	(g*H)(t)=\int_{0}^{\infty} g(t-u)\,du
	=\int_{-\infty}^{t} g(w)\,dw
	=\Phi\!\left(\frac{t}{\sigma}\right),
	\label{eq:9}
\end{equation}
where $\Phi$ is the standard normal cumulative distribution function (CDF)~\cite{CasellaBerger2}
\begin{equation}
	\Phi(x)=\int_{-\infty}^{x}\phi(t)\,dt,
	\qquad
	\phi(t)=\frac{1}{\sqrt{2\pi}}e^{-t^{2}/2}.
	\label{eq:10}
\end{equation}

Inserting Eq.~\eqref{eq:4} into Eq.~\eqref{eq:8}:
\begin{equation}
	S(t;\tau,\sigma)=\int_{0}^{\infty}\frac{1}{\sqrt{2\pi}\sigma}
	\exp\!\left[-\frac{(t-u)^{2}}{2\sigma^{2}}-\frac{u}{\tau}\right]du.
	\label{eq:11}
\end{equation}
Expanding the exponent and defining $\mu\equiv t-\sigma^{2}/\tau$, we obtain
\begin{equation}
	S(t;\tau,\sigma)=
	\exp\!\left(\frac{\sigma^{2}}{2\tau^{2}}-\frac{t}{\tau}\right)
	\int_{0}^{\infty}\frac{1}{\sqrt{2\pi}\sigma}
	\exp\!\left[-\frac{(u-\mu)^{2}}{2\sigma^{2}}\right]du.
	\label{eq:15}
\end{equation}
With the substitution $y=(u-\mu)/\sigma$, the integral yields
\begin{equation}
	\int_{0}^{\infty}\frac{1}{\sqrt{2\pi}\sigma}
	\exp\!\left[-\frac{(u-\mu)^{2}}{2\sigma^{2}}\right]du
	=
	\int_{(-\mu)/\sigma}^{\infty}\phi(y)\,dy
	=
	\Phi\!\left(\frac{\mu}{\sigma}\right).
	\label{eq:16}
\end{equation}
Since $\mu/\sigma=t/\sigma-\sigma/\tau$, we obtain
\begin{equation}
	\boxed{
		S(t;\tau,\sigma)=
		\exp\!\left(\frac{\sigma^{2}}{2\tau^{2}}-\frac{t}{\tau}\right)\,
		\Phi\!\left(\frac{t}{\sigma}-\frac{\sigma}{\tau}\right).
	}
	\label{eq:17}
\end{equation}

Combining Eqs.~\eqref{eq:7}, \eqref{eq:9}, and \eqref{eq:17}:
\begin{equation}
	\boxed{
		F(t;\tau)=1-\Phi\!\left(\frac{t}{\sigma}\right)+
		S(t;\tau,\sigma),
	}
	\label{eq:18}
\end{equation}
where $\sigma$ is related to $\tau_{\mathrm{IRF}}$ by Eq.~\eqref{eq:3}.
For $t\gg \sigma$, $\Phi(t/\sigma)\approx 1$, so $F(t;\tau)\approx S(t;\tau,\sigma)$.

\subsection*{3. Nematic response function and small-anisotropy expansion}

We define
\begin{equation}
	\eta(t)\equiv F(t;\tau_{\parallel})-F(t;\tau_{\perp}),
	\qquad
	\Delta_{\tau}\equiv \tau_{\perp}-\tau_{\parallel},
	\qquad
	\tau\equiv\tau_{\mathrm{avg}}\equiv \frac{\tau_{\parallel}+\tau_{\perp}}{2}.
	\label{eq:19}
\end{equation}
Equivalently, $\tau_{\parallel}=\tau-\Delta_{\tau}/2$ and $\tau_{\perp}=\tau+\Delta_{\tau}/2$. For $\Delta_{\tau}>0$, $\eta(t)$ has a minimum with $\eta_{\min}<0$.
Using Eq.~\eqref{eq:18}, the terms $1-\Phi(t/\sigma)$ cancel out, yielding
\begin{equation}
	\eta(t)=S(t;\tau_{\parallel},\sigma)-S(t;\tau_{\perp},\sigma).
	\label{eq:21}
\end{equation}

For $|\Delta_{\tau}|\ll \tau$, a first-order Taylor expansion of Eq.~\eqref{eq:21} gives
\begin{equation}
	S\!\left(\tau-\frac{\Delta_{\tau}}{2}\right)-S\!\left(\tau+\frac{\Delta_{\tau}}{2}\right)
	\approx -\Delta_{\tau}\,\frac{\partial S}{\partial\tau}(t;\tau,\sigma),
\end{equation}
so
\begin{equation}
	\boxed{
		\eta(t)\approx -\Delta_{\tau}\,\partial_{\tau}S(t;\tau,\sigma).
	}
	\label{eq:22}
\end{equation}
An extremum of $\eta(t)$ satisfies $d\eta/dt=0$. Assuming $\Delta_{\tau}\neq 0$,
\begin{equation}
	\boxed{
		\partial_{t\tau}S(t_{\min};\tau,\sigma)=0.
	}
	\label{eq:23}
\end{equation}

\subsection*{4. Derivation of corrections to $t_{\min}$}

Rewriting Eq.~\eqref{eq:17} as
\begin{equation}
	S(t;\tau,\sigma)=e^{A}\Phi(x),
	\qquad
	A=\frac{\sigma^{2}}{2\tau^{2}}-\frac{t}{\tau},
	\qquad
	x=\frac{t}{\sigma}-\frac{\sigma}{\tau}.
	\label{eq:24}
\end{equation}
Note $\Phi'(x)=\phi(x)$, and $\phi$ is given in Eq.~\eqref{eq:10}.

First derivative with respect to $t$ yields
\begin{align}
	\partial_{t}S
	&=e^{A}\Bigl[(\partial_{t}A)\Phi(x)+\Phi'(x)\,\partial_{t}x\Bigr] \nonumber\\
	&=e^{A}\left[-\frac{\Phi(x)}{\tau}+\frac{\phi(x)}{\sigma}\right],
	\label{eq:25}
\end{align}
since $\partial_{t}A=-1/\tau$ and $\partial_{t}x=1/\sigma$.

We differentiate Eq.~\eqref{eq:25} with respect to $\tau$.
Defining
\begin{equation}
	M\equiv -\frac{\Phi(x)}{\tau}+\frac{\phi(x)}{\sigma},
	\qquad \partial_{t}S=e^{A}M.
	\label{eq:26}
\end{equation}
Then
\begin{equation}
	\partial_{t\tau}S=\partial_{\tau}(e^{A}M)
	=e^{A}\Bigl[(\partial_{\tau}A)M+\partial_{\tau}M\Bigr].
	\label{eq:27}
\end{equation}
Because $e^{A}>0$, Eq.~\eqref{eq:23} is equivalent to
\begin{equation}
	(\partial_{\tau}A)M+\partial_{\tau}M=0.
	\label{eq:28}
\end{equation}
The auxiliary derivatives:
\begin{equation}
	\partial_{\tau}A=-\frac{\sigma^{2}}{\tau^{3}}+\frac{t}{\tau^{2}},
	\qquad
	\partial_{\tau}x=\frac{\sigma}{\tau^{2}}.
	\label{eq:29}
\end{equation}
To compute $\partial_{\tau}M$, we will separately evaluate the derivatives of two terms from Eq.~\eqref{eq:26}. Taking the derivative of the first term,
\begin{equation}
	\partial_{\tau}\!\left(-\frac{\Phi(x)}{\tau}\right)
	=
	\frac{\Phi(x)}{\tau^{2}}
	-\frac{1}{\tau}\phi(x)\,\partial_{\tau}x
	=
	\frac{\Phi(x)}{\tau^{2}}-\frac{\sigma\phi(x)}{\tau^{3}}.
	\label{eq:30}
\end{equation}
For the second term, using $\phi'(x)=-x\phi(x)$,
\begin{equation}
	\partial_{\tau}\!\left(\frac{\phi(x)}{\sigma}\right)
	=
	\frac{1}{\sigma}\phi'(x)\partial_{\tau}x
	=
	\frac{1}{\sigma}(-x\phi(x))\frac{\sigma}{\tau^{2}}
	=
	-\frac{x\phi(x)}{\tau^{2}}.
	\label{eq:31}
\end{equation}
Thus
\begin{equation}
	\partial_{\tau}M=
	\frac{\Phi(x)}{\tau^{2}}
	-\frac{\sigma\phi(x)}{\tau^{3}}
	-\frac{x\phi(x)}{\tau^{2}}.
	\label{eq:32}
\end{equation}

Substituting Eqs.~\eqref{eq:26}, \eqref{eq:29}, and \eqref{eq:32} into Eq.~\eqref{eq:28},
multiplying by $\tau^{3}$, then simplifying using $x=t/\sigma-\sigma/\tau$ yields:
\begin{equation}
	\boxed{
		\left(\tau+\frac{\sigma^{2}}{\tau}-t\right)\Phi(x)=\sigma\,\phi(x),
		\qquad
		x=\frac{t}{\sigma}-\frac{\sigma}{\tau}.
	}
	\label{eq:33}
\end{equation}
Rearranging Eq.~\eqref{eq:33} gives
\begin{equation}
	t=\tau+\frac{\sigma^{2}}{\tau}-\sigma\frac{\phi(x)}{\Phi(x)}
	=\tau+\frac{\sigma^{2}}{\tau}-\sigma\Lambda(x),
	\label{eq:34}
\end{equation}
where $\Lambda(x)$ is the inverse Mills ratio~\cite{Pinelis2019_Mills},
\begin{equation}
	\boxed{
		\Lambda(x)\equiv \frac{\phi(x)}{\Phi(x)}.
	}
	\label{eq:35}
\end{equation}
Therefore,
\begin{equation}
	\boxed{
		t_{\min}=\tau+\frac{\sigma^{2}}{\tau}-\sigma\Lambda(x),
		\qquad
		x=\frac{t_{\min}}{\sigma}-\frac{\sigma}{\tau}.
	}
	\label{eq:36}
\end{equation}

\subsection*{5. Explicit approximation in terms of $\tau_{\mathrm{IRF}}$}

As a first approximation, we neglect the $\sigma\Lambda(x)$ term in Eq.~\eqref{eq:36}:
\begin{equation}
	t_{\min}^{(0)}=\tau+\frac{\sigma^{2}}{\tau}.
	\label{eq:37}
\end{equation}
Inserting Eq.~\eqref{eq:37} into the definition of $x$:
\begin{align}
	x^{(0)}
	&=\frac{t_{\min}^{(0)}}{\sigma}-\frac{\sigma}{\tau}
	=\frac{1}{\sigma}\left(\tau+\frac{\sigma^{2}}{\tau}\right)-\frac{\sigma}{\tau}
	=\frac{\tau}{\sigma}.
	\label{eq:38}
\end{align}
Using Eq.~\eqref{eq:3}, this becomes
\begin{equation}
	\boxed{
		x^{(0)}=\frac{\tau}{\sigma}=\frac{\kappa\,\tau}{\tau_{\mathrm{IRF}}}.
	}
	\label{eq:39}
\end{equation}
Hence an explicit approximation (a one-step fixed-point iteration) is
\begin{equation}
	\boxed{
		t_{\min}\approx \tau+\frac{\sigma^{2}}{\tau}-\sigma\Lambda\!\left(\frac{\tau}{\sigma}\right)
		=
		\tau+\frac{\tau_{\mathrm{IRF}}^{2}}{\kappa^{2}\tau}-\frac{\tau_{\mathrm{IRF}}}{\kappa}\,
		\Lambda\!\left(\frac{\kappa\,\tau}{\tau_{\mathrm{IRF}}}\right).
	}
	\label{eq:40}
\end{equation}

If $u\equiv\tau/\sigma=\kappa\,\tau/\tau_{\mathrm{IRF}}>3$, then $\Phi(u)\approx 1$ and
$\Lambda(u)\approx \phi(u)\propto e^{-u^{2}/2}$ is exponentially small, giving
\begin{equation}
	\boxed{
		t_{\min}\approx \tau+\frac{\sigma^{2}}{\tau}
		=
		\tau+\frac{\tau_{\mathrm{IRF}}^{2}}{\kappa^{2}\tau}
		\qquad
		\left(\frac{\kappa\,\tau}{\tau_{\mathrm{IRF}}}>3\right).
	}
	\label{eq:41}
\end{equation}

For an experimentally relevant case ($\tau_{\mathrm{IRF}}=50\,\mathrm{fs}$ and $\tau\approx 100\,\mathrm{fs}$), the difference between $\tau$ obtained from the full implicit extremum condition (Eq.~\eqref{eq:36}, evaluated numerically) and the estimate based on Eq.~\eqref{eq:41} is below $5\,\mathrm{fs}$ (Fig.~1(f) of the main text). Here $\sigma=\tau_{\mathrm{IRF}}/\kappa\approx 21\,\mathrm{fs}$, so $u\equiv\tau/\sigma\approx 4.7>3$, which satisfies the condition used in Eq.~\eqref{eq:41}.

\subsection*{6. Derivation of the minimum amplitude $\eta_{\min}$ for $u=\kappa\,\tau/\tau_{\mathrm{IRF}}>3$}

For $u>3$ and $t\gtrsim 0$, $\Phi(x)\approx 1$ near the extremum, so Eq.~\eqref{eq:17} simplifies to
\begin{equation}
	S(t;\tau,\sigma)\approx \exp\!\left(\frac{\sigma^{2}}{2\tau^{2}}-\frac{t}{\tau}\right)
	=
	\exp\!\left(\frac{\tau_{\mathrm{IRF}}^{2}}{2\kappa^{2}\tau^{2}}-\frac{t}{\tau}\right).
	\label{eq:42}
\end{equation}

Differentiating Eq.~\eqref{eq:42} with respect to $\tau$:
\begin{align}
	\partial_{\tau}S
	&=S\cdot \partial_{\tau}\!\left(\frac{\sigma^{2}}{2\tau^{2}}-\frac{t}{\tau}\right) \nonumber\\
	&=S\left(-\frac{\sigma^{2}}{\tau^{3}}+\frac{t}{\tau^{2}}\right)
	=S\left(-\frac{\tau_{\mathrm{IRF}}^{2}}{\kappa^{2}\tau^{3}}+\frac{t}{\tau^{2}}\right).
	\label{eq:43}
\end{align}

From Eq.~\eqref{eq:22},
\begin{equation}
	\eta(t)\approx -\Delta_{\tau}\,S(t;\tau,\sigma)
	\left(-\frac{\sigma^{2}}{\tau^{3}}+\frac{t}{\tau^{2}}\right)
	=
	-\Delta_{\tau}\,S(t;\tau,\sigma)
	\left(-\frac{\tau_{\mathrm{IRF}}^{2}}{\kappa^{2}\tau^{3}}+\frac{t}{\tau^{2}}\right).
	\label{eq:44}
\end{equation}

Using Eq.~\eqref{eq:41}, the bracket becomes
\begin{align}
	-\frac{\sigma^{2}}{\tau^{3}}+\frac{t_{\min}}{\tau^{2}}
	&\approx -\frac{\sigma^{2}}{\tau^{3}}
	+\frac{1}{\tau^{2}}\left(\tau+\frac{\sigma^{2}}{\tau}\right) \nonumber\\
	&= -\frac{\sigma^{2}}{\tau^{3}}+\frac{1}{\tau}+\frac{\sigma^{2}}{\tau^{3}}
	=\frac{1}{\tau},
	\label{eq:45}
\end{align}
therefore
\begin{equation}
	\eta_{\min}\approx -\Delta_{\tau}\,\frac{1}{\tau}\,S(t_{\min};\tau,\sigma).
	\label{eq:46}
\end{equation}

Evaluating $S(t_{\min})$ using Eq.~\eqref{eq:42} and $t_{\min}\approx \tau+\sigma^{2}/\tau$:
\begin{align}
	S(t_{\min})
	&\approx \exp\!\left(\frac{\sigma^{2}}{2\tau^{2}}-\frac{1}{\tau}\left(\tau+\frac{\sigma^{2}}{\tau}\right)\right) \nonumber\\
	&=\exp\!\left(\frac{\sigma^{2}}{2\tau^{2}}-1-\frac{\sigma^{2}}{\tau^{2}}\right)
	=e^{-1}\,e^{-\sigma^{2}/(2\tau^{2})} \nonumber\\
	&=e^{-1}\exp\!\left(-\frac{\tau_{\mathrm{IRF}}^{2}}{2\kappa^{2}\tau^{2}}\right).
	\label{eq:47}
\end{align}
Inserting Eq.~\eqref{eq:47} into Eq.~\eqref{eq:46}:

\begin{equation}
	\boxed{
		\eta_{\min}\approx -\,\frac{\Delta_{\tau}}{e\,\tau}\,
		\exp\!\left(-\frac{\tau_{\mathrm{IRF}}^{2}}{2\kappa^{2}\tau^{2}}\right)
		\qquad
		\left(\frac{\kappa\,\tau}{\tau_{\mathrm{IRF}}}>3\right).
	}
	\label{eq:49}
\end{equation}

\subsection*{7. Convolution-corrected results for finite $\tau_{\mathrm{IRF}}$}

For equal amplitudes ($r=1$) and $|\Delta_{\tau}|\ll \tau$:

\begin{equation}
	\boxed{
		t_{\min}\approx \tau+\frac{\tau_{\mathrm{IRF}}^{2}}{\kappa^{2}\tau}
		\qquad
		\left(\frac{\kappa\,\tau}{\tau_{\mathrm{IRF}}}>3\right).
	}
	\label{eq:50}
\end{equation}

From Eq.~\eqref{eq:49},
\begin{equation}
	\boxed{
		\Delta_{\tau}\approx -\,e\,\eta_{\min}\,\tau\,
		\exp\!\left(\frac{\tau_{\mathrm{IRF}}^{2}}{2\kappa^{2}\tau^{2}}\right)
		\qquad
		\left(\frac{\kappa\,\tau}{\tau_{\mathrm{IRF}}}>3\right).
	}
	\label{eq:51}
\end{equation}

Solving Eq.~\eqref{eq:50} for $\tau\equiv\tau_{\mathrm{avg}}$:
\begin{equation}
	t_{\min}\tau=\tau^{2}+\frac{\tau_{\mathrm{IRF}}^{2}}{\kappa^{2}}
	\;\Rightarrow\;
	\tau^{2}-t_{\min}\tau+\frac{\tau_{\mathrm{IRF}}^{2}}{\kappa^{2}}=0,
\end{equation}
so
\begin{equation}
	\boxed{
		\tau\approx
		\frac{t_{\min}+\sqrt{t_{\min}^{2}-4\left(\tau_{\mathrm{IRF}}^{2}/\kappa^{2}\right)}}{2}.
	}
	\label{eq:52}
\end{equation}
For $\kappa\,t_{\min}/\tau_{\mathrm{IRF}}\gg 1$, a simpler approximation is
\begin{equation}
	\boxed{
		\tau\approx t_{\min}-\frac{\tau_{\mathrm{IRF}}^{2}}{\kappa^{2}t_{\min}}=t_{\min}-\frac{\tau_{\mathrm{IRF}}^{2}}{8\ln(2)t_{\min}}.
	}
	\label{eq:53}
\end{equation}

\end{document}